\documentclass[sigconf,camera-ready]{acmart}

\AtBeginDocument{%
  \providecommand\BibTeX{{%
    \normalfont B\kern-0.5em{\scshape i\kern-0.25em b}\kern-0.8em\TeX}}}

\copyrightyear{2020}
\acmYear{2020}
\setcopyright{acmlicensed}\acmConference[CIKM '20]{Proceedings of the 29th ACM International Conference on Information and Knowledge Management}{October 19--23, 2020}{Virtual Event, Ireland}
\acmBooktitle{Proceedings of the 29th ACM International Conference on Information and Knowledge Management (CIKM '20), October 19--23, 2020, Virtual Event, Ireland}
\acmPrice{15.00}
\acmDOI{10.1145/3340531.3412001}
\acmISBN{978-1-4503-6859-9/20/10}
\usepackage{multirow}
\usepackage{pifont}
\usepackage{bm}

\begin{document}
\fancyhead{}
\title{Relational Reflection Entity Alignment}

\author{Xin Mao$^{1*}$, Wenting Wang$^2$, Huimin Xu$^1$, Yuanbin Wu$^1$, Man Lan$^{1*}$}
\email{{xmao,hmxu}@stu.ecnu.edu.cn, {wenting.wang}@lazada.com, {ybwu,mlan}@cs.ecnu.edu.cn}
\affiliation{$^1$East China Normal University, $^2$Alibaba Group}



\begin{abstract}
Entity alignment aims to identify equivalent entity pairs from different Knowledge Graphs (KGs), which is essential in integrating multi-source KGs. 
Recently, with the introduction of GNNs into entity alignment, the architectures of recent models have become more and more complicated.
We even find two counter-intuitive phenomena within these methods:
(1) The standard linear transformation in GNNs is not working well.
(2) Many advanced KG embedding models designed for link prediction task perform poorly in entity alignment.
In this paper, we abstract existing entity alignment methods into a unified framework, \emph{Shape-Builder \& Alignment},
which not only successfully explains the above phenomena but also derives two key criteria for an ideal transformation operation.
Furthermore, we propose a novel GNNs-based method, \emph{Relational Reflection Entity Alignment} (RREA).
RREA leverages \emph{Relational Reflection Transformation} to obtain relation specific embeddings for each entity in a more efficient way.
The experimental results on real-world datasets show that our model significantly outperforms the state-of-the-art methods, exceeding by $5.8\%$-$10.9\%$ on \emph{Hits@1}.
\end{abstract}





 \maketitle

\section{Introduction}
\label{sec:intro}
With more and more KGs emerging, integrating multi-source KGs becomes necessary and beneficial to not only complement information but also improve downstream tasks such as recommendation system and search engine.
One of the key steps to integrating KGs is to identify equivalent entity pairs.
Therefore, the task of entity alignment attracts increasing attention in recent years.
Existing entity alignment methods can be divided into two main categories:
(1) \textbf{Translation-based.}
Inspired by cross-lingual word embedding task, these methods presume that embeddings of different KGs have similar distributions, so the entity pairs who are aligned between KGs would also have relatively similar positions in their own vector spaces.
These methods \cite{DBLP:conf/ijcai/ChenTYZ17,DBLP:conf/semweb/SunHL17,DBLP:conf/ijcai/SunHZQ18,DBLP:conf/icml/GuoSH19} first use translation-based KGs embedding models (e.g., TransE \cite{DBLP:conf/nips/BordesUGWY13}) on every single KG to get its embeddings of entities and relations, and then align entities from two vector spaces into a unified one based on some pre-aligned entity pairs.
(2) \textbf{GNNs-based.}
Different from translation-based methods where the relation is a translation from one entity to another, Graph Neural Networks (GNNs) generate node-level embeddings through aggregating information from the neighboring nodes.
Inspired by Siamese Neural Networks \cite{DBLP:conf/cvpr/ChopraHL05} which are widely used in computer vision, a typical architecture of GNNs-based methods \cite{DBLP:conf/acl/CaoLLLLC19,sun2019knowledge,DBLP:conf/emnlp/WangLLZ18} consists of two multi-layer GNNs with the contrastive loss \cite{DBLP:conf/cvpr/HadsellCL06} or triplet loss \cite{DBLP:conf/cvpr/SchroffKP15}.

With the introduction of GNNs into entity alignment task, recent model architectures have become more and more complicated which are hard to interpret the effectiveness of individual components.
Despite the success in empirical results, we observe two counter-intuitive phenomena in these complicated methods that need to be further clarified and studied:

Q$1$: \textbf{Why the standard linear transformation of GNNs is not working well in entity alignment?}
GNNs are originally designed with a standard linear transformation matrix,
however, many GNNs-based methods \cite{DBLP:conf/emnlp/WangLLZ18,DBLP:conf/acl/CaoLLLLC19,li-etal-2019-semi,yang2019aligning}\footnote{GCN-Align:\url{https://github.com/1049451037/GCN-Align}; MuGNN: \url{https://github.com/thunlp/MuGNN}; HMAN: \url{https://github.com/h324yang/HMAN}; KECG: \url{https://github.com/THU-KEG/KECG}}
constrain it to be unit (i.e., removing this matrix from GNNs) or diagonal with unit initialization.
All previous methods just treat it as parameter reduction but do not explore nor explain about this setting.
When we try to undo this setting in GCN-Align \cite{DBLP:conf/emnlp/WangLLZ18}, the performances significantly drop by $\geqslant10\%$ on \emph{Hits@1}.
So we believe this should be related to some more fundamental issues.

Q$2$: \textbf{Why many advanced KG embedding models are not working well in entity alignment?}
In other tasks that also need KG modeling, such as link prediction, many advanced KG embedding models are proposed and proved to be very effective.
Strangely, a lot of these advanced embedding models designed for link prediction do not show success in entity alignment.
\citet{DBLP:conf/semweb/SunHHCGQ19} experiments with many advanced KG embedding models, such as TransR \cite{DBLP:conf/aaai/LinLSLZ15}, ConvE \cite{DBLP:conf/aaai/DettmersMS018} and etc., but performances are even worse than TransE.
The authors conclude with "not all embedding models designed for link prediction are suitable for entity alignment" but not giving any further exploration or explanation.

To analyze these two issues from a global and unified perspective, we propose an abstract entity alignment framework, named as \emph{Shape-Builder \& Alignment}.
In this framework, both translation-based and GNNs-based methods are just special cases under respective special settings.
With this framework, we successfully derive the answers to address the above questions:
(Q$1$) Entity alignment presumes similarity between distributions, so in order to avoid destroying the shape, the norms and the relative distances of entities should remain unchanged after transformation.
Thus, it is mandatory that the transformation matrix is orthogonal.
(Q$2$) Many advanced KG embedding models share one key idea --- transforming  entity embeddings into relation specific ones.
However, their transformation matrix is difficult to comply with the orthogonal property.
This is the fundamental reason why they perform poorly in entity alignment.

Inspired by the above findings, we propose two key criteria of an ideal transformation operation for entity alignment: \textbf{Relational Differentiation} and \textbf{Dimensional Isometry}.
Then, we design a new transformation operation, \emph{Relational Reflection Transformation}, which fulfills these two criteria.
This new operation is able to reflect entity embeddings along different relational hyperplanes to construct relation specific embeddings.
Meanwhile, the reflection matrix is orthogonal which is easy to prove, so reflection transformation could keep the norms and the relative distances unchanged.
By integrating this proposed transformation into GNNs, we further present a novel GNNs-based entity alignment method, \emph{Relational Reflection Entity Alignment} (RREA).
The experimental results on real-world public datasets validate that our model greatly exceeds existing state-of-the-art methods by $5.8\%$-$10.9\%$ on \emph{Hits@1} across all datasets.
We summarize the main contributions of this paper as follows:
\begin{itemize}
  \item To our best knowledge, this is the first work to abstract existing entity alignment methods into a unified framework.
      Through this framework, we successfully derive two key criteria for an ideal transformation operation: relational differentiation and dimensional isometry.
  \item To our best knowledge, this is the first work to design a new transformation operation, \emph{Relational Reflection Transformation}, which fulfills the above two criteria.
        By integrating this operation into GNNs, we further propose a novel GNNs-based method \emph{Relational Reflection Entity Alignment} (RREA).
  \item The extensive experimental results show that our model is ranked consistently as the best across all real-world datasets and outperforms the state-of-the-art methods by $5.8\%$-$10.9\%$ on \emph{Hits@1}.
      In addition, we also carry ablation experiments to demonstrate that each component of our model is effective.

\end{itemize}

\section{Related Work}
\label{sec:RW}
Existing entity alignment methods can be divided into two categories according to their motivations.
In this section, we will give a detailed illustration of these methods.

\subsection{Translation-based Methods}
\begin{figure}
 \resizebox{0.95\linewidth}{!}{
  \centering
  \includegraphics[width=\linewidth]{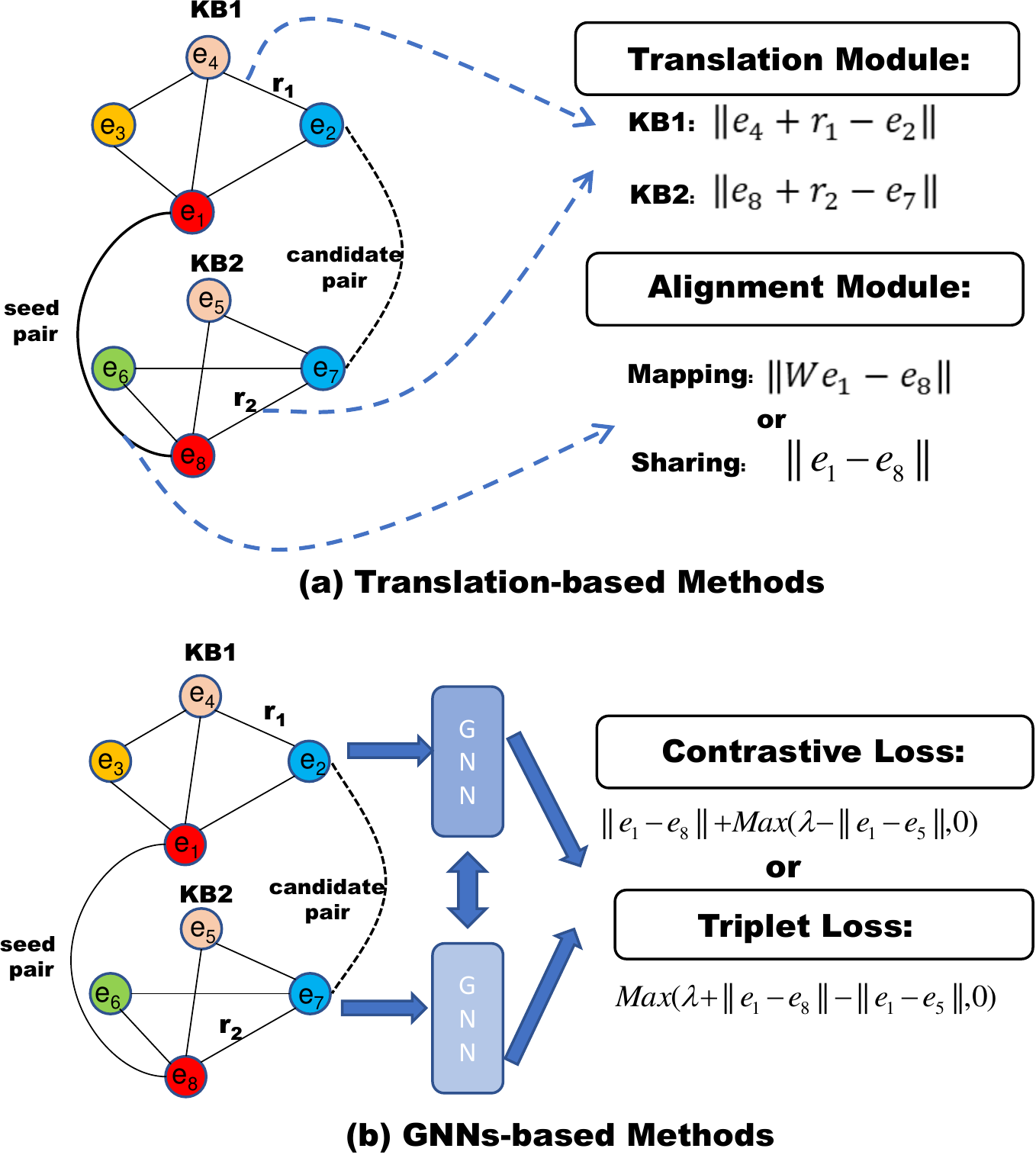}
  }
  \caption{Decomposition of existing alignment methods.}\label{figure:rw}
\end{figure}

Translation-based methods are originated from cross-lingual word embedding task.
So they also have a core assumption that the entity embeddings of different KGs have similar distributions, just like the word embeddings of different languages.
As shown in Figure \ref{figure:rw}(a), translation-based methods usually consist of two modules: translation module and alignment module.

\textbf{Translation Module}:
The major function of the translation module is to constrain the randomly initialized embeddings into a fixed distribution through translation-based KGs embedding models.
Due to its solid theoretical foundation and minimum implementation effort, the majority of translation-based methods adopt TransE \cite{DBLP:conf/nips/BordesUGWY13} as the translation module (e.g., MtransE \cite{DBLP:conf/ijcai/ChenTYZ17}, JAPE \cite{DBLP:conf/semweb/SunHL17} and BootEA \cite{DBLP:conf/ijcai/SunHZQ18}).
Inspired by Word$2$Vec \cite{DBLP:journals/corr/MikolovLS13}, TransE interprets a relation as the translation from its head to its tail ($\bm{h} + \bm{r} \approx \bm{t}$), so that entity embeddings also have the property of translation invariance.
Theoretically, any KG embedding model could act as a translation module.
However, as mentioned in Section \ref{sec:intro}, many advanced embedding models \cite{DBLP:conf/aaai/DettmersMS018,DBLP:conf/aaai/LinLSLZ15} which perform well in link prediction do not show success in entity alignment.

\textbf{Alignment Module}:
By taking pre-aligned entities as seeds, the alignment module is responsible for aligning the embeddings of different KGs into a unified vector space.
At present, there are two types of alignment modules:

($1$) \emph{mapping}:
Similar to its counterparts in cross-lingual word embedding, this approach embeds different KGs into a unified vector space through a linear transformation matrix.
For example, MtransE \cite{DBLP:conf/ijcai/ChenTYZ17}, KDCoE \cite{DBLP:conf/ijcai/ChenTCSZ18}, and OTEA \cite{DBLP:conf/ijcai/Pei0Z19} minimize the distances between the pre-aligned pairs by optimizing one or two linear transformation matrices (i.e., $ {\bm{W}\bm{e}_1\approx \bm{e}_2}$ or $ {\bm{W}_1\bm{e}_1\approx \bm{W}_2\bm{e}_2}$).

($2$) \emph{sharing}:
The \emph{sharing} approach embeds different KGs into a unified vector space by letting each pre-aligned pair directly share the same embedding, which is more straightforward compared to the \emph{mapping} approaches.
There are three different implementations about \emph{sharing}:
(a) MTransE \cite{DBLP:conf/ijcai/ChenTYZ17} proposes to minimize the equation ${\left\| \bm{e}_1 - \bm{e}_2\right\|}$ for each pre-aligned pairs\footnote{Hereafter, $\|*\|$ means L$1$ or L$2$ norm unless explicitly specified.}.
(b) JAPE \cite{DBLP:conf/semweb/SunHL17} and RSNs \cite{DBLP:conf/icml/GuoSH19} directly configure ${e_1}$ and $\rm{e_2}$ to share a common embedding when the model is built.
(c) BootEA \cite{DBLP:conf/ijcai/SunHZQ18} and TransEdge \cite{DBLP:conf/semweb/SunHHCGQ19} swap the pre-aligned entities in their triples to generate extra triples for supervision,
e.g., given $(e_1, e_2)$ is a pre-aligned pair and a triple $\langle e_1, r_1, e_3\rangle$ in KGs, the model will produce a new triple $\langle e_2, r_1, e_3\rangle$.

\subsection{GNNs-based Methods}
Due to the fact that TransE is only trained on individual triples, it may lack the ability to exploit the global view of entities and relations.
Therefore, many recent studies introduce GNNs into entity alignment task, which is originated with the ability to model global information of graphs.

Inspired by Siamese Neural Networks \cite{DBLP:conf/cvpr/ChopraHL05}, a typical GNNs-based method has a simple and intuitive architecture (as shown in Figure \ref{figure:rw}(b)) --- two multi-layer GNNs encoders with a loss function, either contrastive loss \cite{DBLP:conf/cvpr/HadsellCL06} or triplet loss \cite{DBLP:conf/cvpr/SchroffKP15}.
The first GNNs-based method is proposed by GCN-Align \cite{DBLP:conf/emnlp/WangLLZ18} using multi-layer vanilla GCN as the encoder and successfully applies GNNs to entity alignment task.
However, due to the disability of vanilla GCN in modeling heterogeneous graphs, GCN-Align is unable to effectively utilize the rich relation information in KGs.

Many more recent studies attempt to incorporate relation information into GNNs and build relation-aware models to better represent KGs.
HMAN \cite{yang2019aligning} concatenates the entity embeddings obtained by GCN with the average of the neighboring relation and attribute embeddings.
MuGNN \cite{DBLP:conf/acl/CaoLLLLC19}, NAEA \cite{DBLP:conf/ijcai/ZhuZ0TG19} and MRAEA \cite{DBLP:conf/wsdm/MaoWXLW20} assign different weight coefficients to entities according to relation types between them, which empowers the model to distinguish the importance between different entities.
RDGCN \cite{DBLP:conf/ijcai/WuLF0Y019} establishes a dual relation graph for KGs which regards relation as node and entity as edge.
Strangely, many GNNs-methods \cite{DBLP:conf/acl/CaoLLLLC19,DBLP:conf/emnlp/WangLLZ18,li-etal-2019-semi,yang2019aligning,DBLP:conf/wsdm/MaoWXLW20} adopt counter-intuitive constraint in their transformation matrix design,
i.e., forcing the matrix to be unit or diagonal.
All previous methods just treat it as parameter reduction but do not explore nor explain about this setting.

In addition, there are also some other GNNs-based models proposed for modeling KGs in link prediction task.
By assigning different transformation matrices to different relations, RGCN \cite{DBLP:conf/esws/SchlichtkrullKB18} maps entities to corresponding relational vector spaces before convolution.
KBAT \cite{DBLP:conf/acl/NathaniCSK19} converts the triple embeddings into new entity embeddings with a linear transformation matrix and assigns different weight coefficients to the new embeddings via attention mechanism.
However, according to our experimental results in Table \ref{table:KEM}, these advanced models perform even worse than vanilla GCN in entity alignment.

\section{Preliminary}
\subsection{Problem Formulation}
KGs store the real-world information in the form of triples, $\langle entity_1,$
$relation, entity_2\rangle$, which describe the relations between two entities.
A KG could be defined as $G=(E,R,T)$, where $E$ and $R$ represent the sets of entities and relations respectively, $T$ represents the set of triples.
Although different KGs are constructed from different sources, there are still many entity pairs referring to the same real-world object.
Entity alignment aims to find these aligned entity pairs from multi-source KGs, which is the key step of knowledge integration.
Formally, $G_1$ and $G_2$ are two multi-source KGs, $P=\left\{(e_{i_1},e_{i_2})|e_{i_1}\in E_1,e_{i_2}\in E_2\right\}^p_{i=1}$ represents the set of pre-aligned seed pairs.
The aim of entity alignment is to find new aligned entity pairs based on these pre-aligned seeds.

\begin{table}[t]
\begin{center}
\resizebox{0.95\linewidth}{!}{
\begin{tabular}{cc|cccc}
\hline
\multicolumn{2}{c|}{Datasets} & Entity & Relation  & Triple \\
\hline
\multirow{2}{1.3cm}{$\rm{DWY_{YG}}$} & DBpedia & 100,000 &  302  & 428,952  \\
& YAGO3 & 100,000 & 31  &  502,563 \\
\hline
\multirow{2}{1.3cm}{$\rm{DWY_{WD}}$} & DBpedia & 100,000 &  330 & 463,294 \\
& Wikipedia & 100,000 &  220  &  448,774  \\
\hline
\multirow{2}{1.3cm}{$\rm{DBP_{ZH-EN}}$} & Chinese & 66,469 & 2,830& 153,929\\
& English & 98,125 & 2,317 & 237,674 \\
\hline
\multirow{2}{1.3cm}{$\rm{DBP_{JA-EN}}$} & Japanese & 65,744 & 2,043 & 164,373 \\
& English & 95,680 & 2,096  & 233,319 \\
\hline
\multirow{2}{1.3cm}{$\rm{DBP_{FR-EN}}$} & French & 66,858 & 1,379 & 192,191 \\
& English & 105,889 & 2,209 & 278,590  \\
\hline
\end{tabular}
}
\end{center}
\caption{Statistical data of DBP15K and DWY100K.}\label{table:data1}
\end{table}

\subsection{Datasets}
In order to make the comparison with previous methods reliable and fair, we experiment on two widely used open-source datasets:
\begin{itemize}
  \item $\rm{DBP15K}$ \cite{DBLP:conf/semweb/SunHL17} which contains three cross-lingual datasets constructed from the multilingual version of DBpedia, including $\rm{DBP_{ZH-EN}}$ (Chinese to English), $\rm{DBP_{JA-EN}}$ (Japanese to English), and $\rm{DBP_{FR-EN}}$ (French to English).
  \item $\rm{DWY100K}$ \cite{DBLP:conf/ijcai/SunHZQ18} are extracted from DBpedia, Wikidata, and YAGO$3$. It has two monolingual datasets: $\rm{DWY_{WD}}$ (DBpedia-Wikidata) and $\rm{DWY_{YG}}$ (DBpedia-YAGO$3$). Each dataset has $100,000$ reference entity alignments and more than nine hundred thousand triples.
\end{itemize}
Table \ref{table:data1} shows the statistics of these datasets.
Following the setting of previous studies, we randomly split $30\%$ of aligned pairs for training and keep $70\%$ of them for testing.
The reported performance is the average of five independent training runs and the train/test datasets are shuffled in every round.

\section{A Unified Entity Alignment Framework}
\label{sec:transformation}
In this section, we model GNNs-based methods and translation-based methods into an abstract but unified entity alignment framework.
Then this framework successfully leads to not only the answers regarding the two questions raised in Section \ref{sec:intro} but also the key criteria of an ideal transformation operation for entity alignment.

\subsection{Shape-Builder \& Alignment}
\label{4.1}
The motivation behind translation-based entity alignment methods is cross-lingual word embedding (word alignment).
So naturally, they all can be abstracted into a unified framework composed of \textbf{Shape-Builder} and \textbf{Alignment} as shown in Figure \ref{fig:1}:
\begin{figure}[t]
  \centering
  \includegraphics[width=\linewidth]{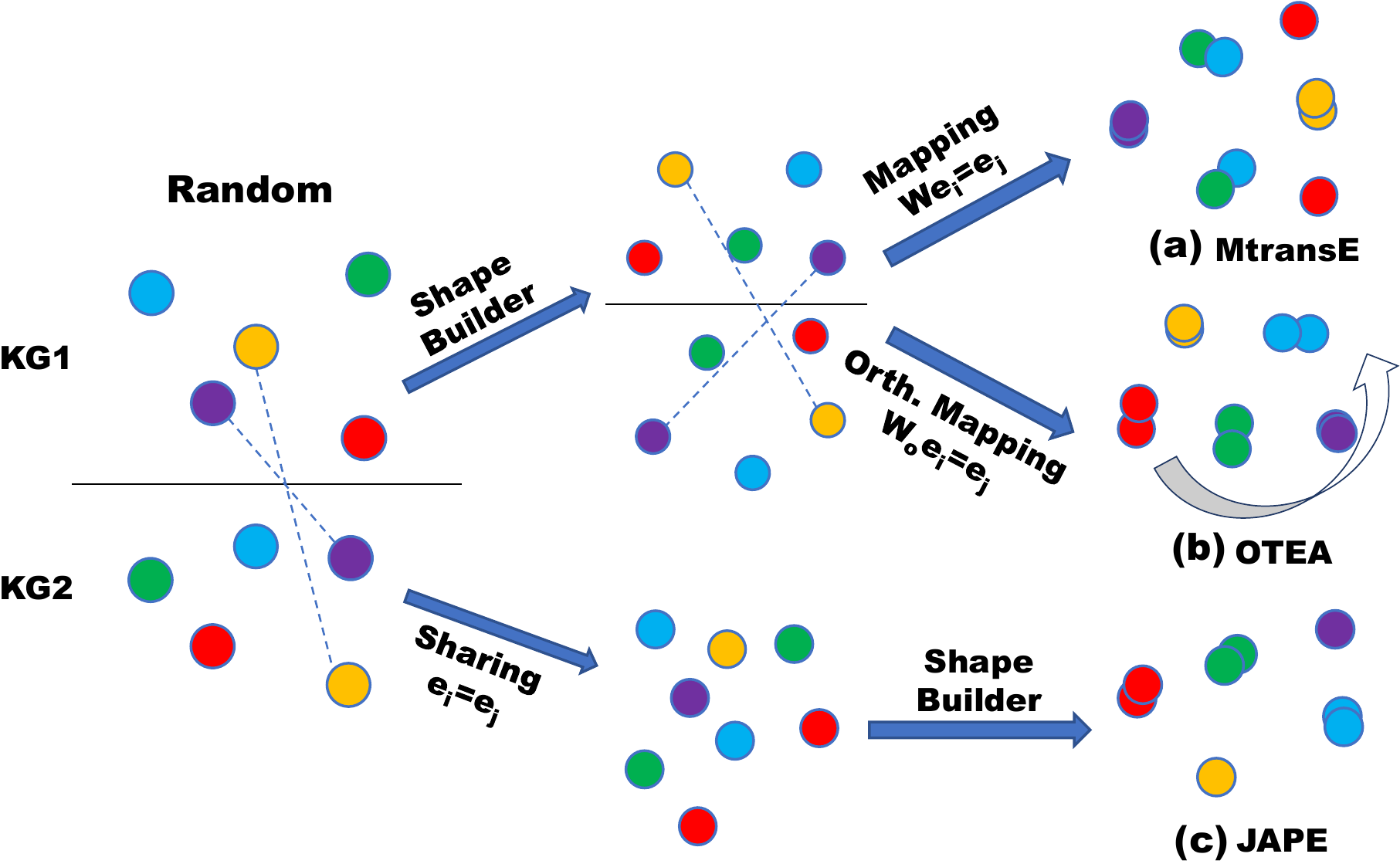}\\
  \caption{The unified framework of entity alignment and representative alignment methods.
}\label{fig:1}
\end{figure}

\textbf{Shape-Builder}:
The main function of shape-builder is to constrain the random initialized distribution to a specific distribution which we define as \emph{shape}.
Obviously, the \emph{translation module} mentioned in Section \ref{sec:RW} is a shape-builder.
In fact, besides TransE, any embedding model can be used as a shape-builder.
The only prerequisite is that the obtained embeddings from two KGs should have \emph{Shape Similarity} in-between. 
In other words, equivalent elements (such as word or entity) have relatively similar positions in their own vector spaces.

\textbf{Alignment}:
When the \emph{Shape Similarity} holds, different shapes can be matched by pre-aligned seeds.
As mentioned in Section \ref{sec:RW}, \emph{mapping} is one of the alignment modules in translation-based methods which trains a matrix $\bm{W}$ to minimize the distances between the pre-aligned seeds \cite{DBLP:conf/ijcai/ChenTYZ17} as follow:
\begin{equation}
  \underset W{min\;}\sum_{(e_i,e_j)\in P}\left\|\bm{W}\bm{h}_{e_i}-\bm{h}_{e_j}\right\|
\end{equation}
where $(e_i,e_j)$ is a pre-aligned pair, $\bm{h}_{e_i}$ represents the embedding vector of entity $e_i$.
However, if matrix $\bm{W}$ has no constraint, then there is no guarantee that the norms and the relative distances of embeddings will be reserved after transformation, which in turn could destroy the original shape similarity.
The seed pairs are well fitted, but the rest of entities could be misaligned (as illustrated in Figure \ref{fig:1}(a)).
On the other hand, if $\bm{W}$ is constrained to be orthogonal, it becomes a rotation operation and then shape similarity will not be destroyed.
This is why many word alignment methods\cite{DBLP:conf/naacl/XingWLL15,DBLP:conf/iclr/SmithTHH17} use orthogonal constraint.
In entity alignment, OTEA \cite{DBLP:conf/ijcai/Pei0Z19} also proposes to constrain the transformation matrix to be orthogonal (as illustrated in Figure \ref{fig:1}(b)).

In addition, in another alignment module \emph{sharing}, pre-aligned entities are treated as anchors and then the rest of the entities can be gradually aligned during the optimization process of shape-builder (as illustrated in Figure \ref{fig:1}(c)).
Compared to \emph{mapping}, \emph{sharing} abandons the transformation matrix at all which reduces parameters and simplifies the architecture.
So far, all translation-based methods could be abstracted into this framework.

\subsection{GNNs-based Methods Are Also Subject to Our Unified Framework}
\label{4.2}
Many GNNs in entity alignment task contains the following equa-\\tions\cite{DBLP:conf/nips/HamiltonYL17}:
\begin{equation}
h^{l}_{{\mathcal N}_{e_i}^e} \leftarrow {Aggregate}(\{\bm{h}_{e_k}^{l}, \forall e_k \in \{e_i\} \cup {\mathcal N}_{e_i}^e\})
\label{eq2}
\end{equation}
\begin{equation}
h^{l+1}_{e_i} \leftarrow \sigma\left(\bm{W}^l\cdot h^{l}_{\mathcal N_{e_i}^e}\right)
\label{eq3}
\end{equation}
where ${\mathcal N}_{e_i}^e$ represents the set of neighboring nodes around $e_i$, $\bm{W}^l$ is the transformation matrix of layer $l$.
Equation \ref{eq2} is responsible for aggregating information from the neighboring nodes while Equation \ref{eq3} transforms the node embeddings into better ones.
There are many operations available that can serve the purpose of $Aggregate$, such as normalized mean pooling (vanilla GCN\cite{DBLP:conf/iclr/KipfW17}) and attentional weighted summation (GAT \cite{DBLP:conf/iclr/VelickovicCCRLB18}).

After generating the embeddings, GNNs-based methods often use triplet loss to make the equivalent entities close to each other:
\begin{equation}
L=\sum_{\substack{(e_i,e_j)\in P\\(e_i',e_j')\in P'}}max\left(\underset{alignment}{\underline{\|\bm{h}_{e_i}-\bm{h}_{e_j}\|}}-\underset{apart}{\underline{\|\bm{h}_{e_i'}-\bm{h}_{e_j'}\|+\lambda}},0\right)
\end{equation}
where $\lambda$ represents the margin hyper-parameter, $(e_i',e_j')$ represents the negative pair by randomly replacing one of $(e_i,e_j)$.
Interestingly, the first half of the loss function (i.e., $\|\bm{h}_{e_i}-\bm{h}_{e_j}\|$) is exactly the same as the \emph{sharing} alignment module.
The same finding is even more obvious if looking at the contrastive loss used in AliNet\cite{sun2019knowledge}:
\begin{equation}
L=\sum_{(e_i,e_j)\in P}\underset{alignment}{\underline{\|\bm{h}_{e_i}-\bm{h}_{e_j}\|}}+\sum_{(e_i',e_j')\in P'}max\left(\underset{apart}{\underline{\|\bm{h}_{e_i'}-\bm{h}_{e_j'}\|+\lambda}},0\right)
\end{equation}
So the losses in GNNs all can be broken down into two sub-parts: the $1st$ half, i.e. alignment loss, acts as an alignment module; while the $2nd$ half, i.e. apart loss, acts as part of a shape-builder.

Therefore, we propose a hypothesis: \textbf{GNNs-based methods are also subject to our unified framework, \emph{Shape-Builder \& Alignment}}.
More specifically, we believe the $Aggregate$ operation of GNNs and the apart loss function together compose a potential shape-builder.
The $Aggregate$ operation makes similar entities close to each other, and the apart loss keeps dissimilar entities away from each other.
So the combination of them builds a distribution which possess the property of \emph{Shape Similarity}.

\begin{figure}[t]
  \centering
  \includegraphics[width=\linewidth]{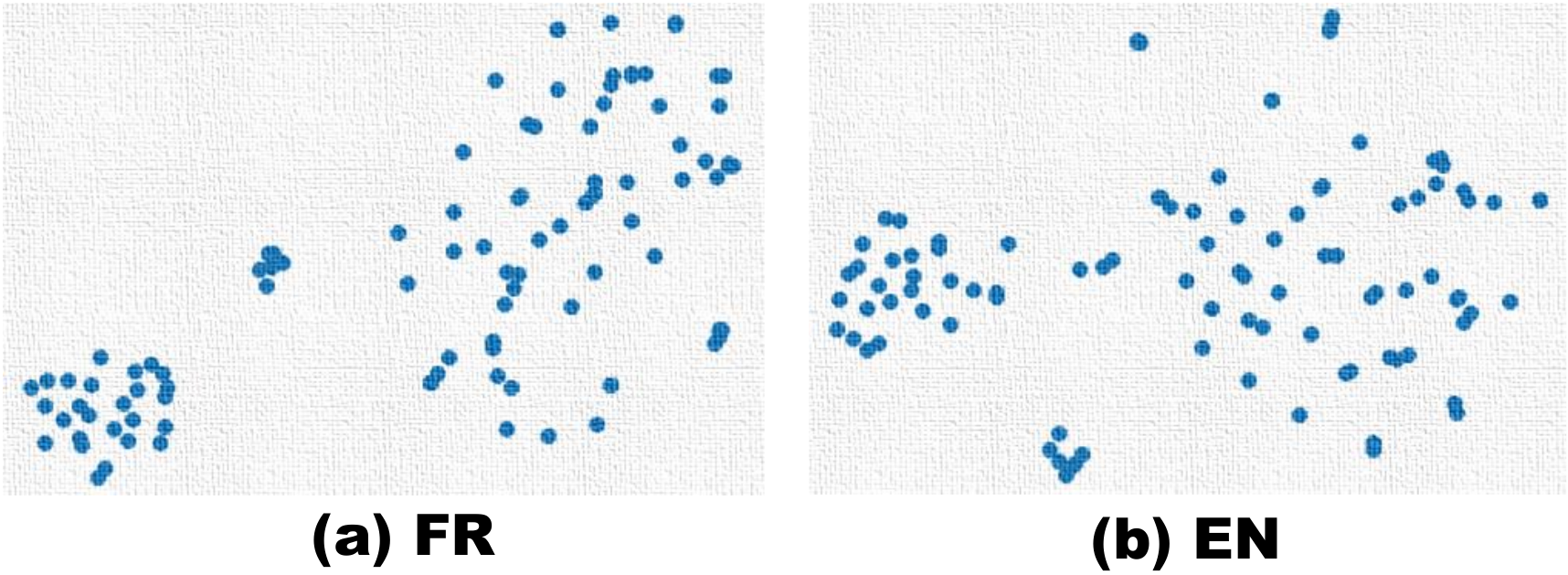}
  \caption{The distributions obtained by GCN-Align on $\bm{DBP_{FR-EN}}$.}\label{distribution}
\end{figure}

\noindent
\textbf{Visual Experiment}:
If our hypothesis is correct, distributions of different KGs should have visual similarity.
Thus, to verify our hypothesis, we retain the apart loss from triplet loss in GCN-Align \cite{DBLP:conf/emnlp/WangLLZ18}\footnote{Although \citet{DBLP:conf/emnlp/WangLLZ18} retain $W$ in the paper, it is actually removed from the released code.} which has the simplest architecture:
\begin{equation}
  L_{apart} = \sum_{(e_i',e_j')\in P'}max\left(\lambda-{\left\|\bm{h}_{e_i{'}}-\bm{h}_{e_j{'}}\right\|}_1,0\right)
\end{equation}
Then GCN-Align is transformed from a supervised model into a self-supervised model.
We train the model on $\rm DBP_{FR-EN}$ and extract $100$ embeddings of aligned pairs, then map them to $2$-dimensional space by t-SNE \cite{Hinton2008Visualizing}.
The distributions are shown in Figure \ref{distribution} and we observe that there indeed are similarities between the two distributions.
For instance, both of them have a large amount of entities scattered in the right portion while having a small amount of entities located closely in the left bottom corner.

\noindent
\textbf{Quantitative Experiment}:
If the distributions have \emph{shape similarity}, the relative distances between entities in one KG should be equal to that of the counterparts in another KGs.
To further quantify the similarity between the two distributions, we design shape similarity metric as follows:
\begin{equation}
  {\rm \emph{SS}}=\frac{\sum_{(e_i,\widetilde{e_i})\in P}\sum_{(e_j,\widetilde{e_j})\in P}dist(e_i,e_j)\;-\;dist(\widetilde{e_i},\widetilde{e_j})}{\sum_{(e_i',\widetilde{e_i}')\in P'}\sum_{(e_j',\widetilde{e_j}')\in P'}dist(e_i',e_j')\;-\;dist(\widetilde{e_i}',\widetilde{e_j}')}
\label{eq7}
\end{equation}
where $e_i,e_j \in G_1$ represent an arbitrarily entity pair in one KG and $\widetilde{e_i},\widetilde{e_j} \in G_2$ represent the counterparts in another KG.
Then $(e_i',\widetilde{e_i}',e_j',\widetilde{e_j}')$ represents a negative quadruple obtained by randomly replacing one entity from $(e_i,\widetilde{e_i},e_j,\widetilde{e_j})$,
$dist(e_i,e_j)$ represents the distance between two entities where any distance metrics such as L$2$ or cosine is applicable.
All the embeddings are normalized by L$2$-normalization.
In Equation \ref{eq7}, the numerator represents the difference of distances between aligned entities, while the denominator represents that of random pairs.

\begin{table}[t]
\renewcommand\arraystretch{1.2}
\centering
\resizebox{1.0\linewidth}{!}{
\begin{tabular}{c|cc|cc|cc}
\hline
\multirow{2}{* }{Method} & \multicolumn{2}{c|}{$\rm{DBP_{ZH-EN}}$} & \multicolumn{2}{c|}{$\rm{DBP_{JA-EN}}$} & \multicolumn{2}{c}{$\rm{DBP_{FR-EN}}$}\\
& Cosine & L2 & Cosine & L2 & Cosine & L2\\
\hline
 Random&0.997&0.996&0.996&0.995&0.998&0.997\\
 TransE&0.682&0.657&0.693&0.665&0.548& 0.534\\
 GCN-Align(w/o T)& 0.982&0.976&0.979&0.981&0.975&0.968\\
 GCN-Align& \textbf{0.664}&\textbf{0.641}&\textbf{0.676}&\textbf{0.643}&\textbf{0.526}&\textbf{0.498}\\
\hline
\end{tabular}
}
\caption{The \emph{SS} on different datasets and metrics. "w/o T" represents without training.}
\label{table:diff}
\end{table}

Ideally, the \emph{SS} between the distributions should be as small as possible and the \emph{SS} between the random distributions should be close to$1$.
Table \ref{table:diff} shows the \emph{SS} between the distributions obtained by random initialization, GCN-Align, and TransE under two different distance metrics.
The experimental results are in line with our expectation:
(1) The \emph{SS} between the random embeddings is almost $1$.
(2) Although the untrained GCN-Align has some minimum clustering ability, it is still close to the random initialization.
(3) Both TransE and GCN-Align successfully reduce the \emph{SS} of the distributions and GCN-Align is slightly better than TransE.

These two experiments prove that the \emph{Aggregate} operation of GNNs and the apart loss compose a shape-builder together.
Notice that our hypothesis is applicable to the alignment methods purely based on structural information (i.e., triples).
Some methods \cite{DBLP:conf/acl/XuWYFSWY19,DBLP:conf/emnlp/WuLFWZ19} take entity names and pre-align them by machine translation or cross-lingual word embeddings.
In these methods, GNNs play a role as noise smoothing rather than actual alignment.
Therefore, these methods are not in the scope of our framework.

\subsection{Why Linear Transformation Not Work}
\label{4.3}
As mentioned in Section \ref{sec:intro}, many GNNs-based methods \cite{DBLP:conf/acl/CaoLLLLC19,yang2019aligning,li-etal-2019-semi,DBLP:conf/wsdm/MaoWXLW20} constrain their transformation matrix to be unit (i.e., removing $W$) or diagonal with unit initialization.
With our hypothesis verified in Section \ref{4.2}, it is easy to explain why these methods adopt such a counter-intuitive constraint.
In fact, if transformation matrix $\bm{W}$ of GNNs is a unit matrix, it is equivalent to $sharing$ alignment in translation-based methods;
If $\bm{W}$ is unconstrained, it is equivalent to $mapping$ alignment in translation-based methods.
As explained in Section \ref{4.1}, the unconstrained transformation could destroy \emph{Shape Similarity} and degrade performances.
Therefore, the orthogonal constraint should be adopted to reserve the norm and relative distance during transformation.
In fact, unit matrix is not only a special case of orthogonal but also the simplest implementation.
In order to verify our answer to Q$1$, we design two experiments:

\noindent
(1) \textbf{Experiment on GCN-Align}:
To prove that keeping transformation matrix orthogonal is necessary,
we test different constraints on GCN-Align which is the simplest GNN-based method\footnote{In our experiment, dropout  rate is set to $30\%$. There's no dropout in original code of GCN-Align, so our experiment results are higher than that in origin paper.}.
To keep $\bm{W}$ orthogonal in the training process, we adopt the following constraint:
\begin{equation}
L_o=\left\|\bm{W}^{T}\bm{W}-\bm{I}\right\|^2_2
\end{equation}
From Table \ref{table:1},
it's not surprising to see that the unconstrained method is the worst.
Although diagonal constraint with unit initialization shows a great improvement, both unit and orthogonal $W$ achieve the best and very close performances.
This indicates that diagonal constraint is only a temporary solution under incomplete understanding.
Orthogonal initialization with unconstrained $\bm{W}$ slightly improves the performance compared to He initialization,
but the large gap between unconstrained $\bm{W}$ and orthogonal $\bm{W}$ demonstrates that orthogonal constraint is an essential factor impacting performance.

\begin{table}[t]
\renewcommand\arraystretch{1.2}
\centering
\resizebox{1.0\linewidth}{!}{
\begin{tabular}{c|cc|cc|cc}
\hline
\multirow{2}{* }{Method} & \multicolumn{2}{c|}{$\rm{DBP_{ZH-EN}}$} & \multicolumn{2}{c|}{$\rm{DBP_{JA-EN}}$} & \multicolumn{2}{c}{$\rm{DBP_{FR-EN}}$}\\
& Hits@1 & MRR & Hits@1 & MRR & Hits@1 & MRR\\
\hline
 Unconst. $W$ (He Init.\cite{DBLP:conf/iccv/HeZRS15}) &0.340&0.465&0.361&0.483&0.344& 0.481\\
 Unconst. $W$ (Orth. Init.)& 0.349&0.475&0.374&0.496&0.351&0.485 \\
 Diagonal $W$ (Unit Init.)&0.438&0.563&0.449&0.573&0.453& 0.589\\
 Unit $W$&\textbf{0.449}&\textbf{0.574}&0.464&0.588&\textbf{0.463}&\textbf{0.596} \\
 Orthogonal $W$&0.448&0.573&\textbf{0.466}&\textbf{0.589}&0.462&0.594\\
\hline
\end{tabular}
}
\caption{Performances on DBP15K with different constraints and initializations. "Unconst." represents unconstrained. "Orth. Init." represents orthogonal initialization.}
\label{table:1}
\end{table}

\begin{table}[t]
\renewcommand\arraystretch{1.2}
\centering
\resizebox{1.0\linewidth}{!}{
\begin{tabular}{c|cc|cc|cc}
\hline
\multirow{2}{* }{Method} & \multicolumn{2}{c|}{$\rm{DBP_{ZH-EN}}$} & \multicolumn{2}{c|}{$\rm{DBP_{JA-EN}}$} & \multicolumn{2}{c}{$\rm{DBP_{FR-EN}}$}\\
& Hits@1 & MRR & Hits@1 & MRR & Hits@1 & MRR\\
\hline
 MuGNN &0.494&0.611&0.501&0.621&0.495 &0.621 \\
 MuGNN (Orth. $W$)& 0.502& 0.614&0.508 &0.623& 0.511 & 0.627\\
 MuGNN (Unit $W$)& \textbf{0.505}&\textbf{0.617}&\textbf{0.511}&\textbf{0.629}&\textbf{0.514} &\textbf{0.637}\\
 \hline
 KECG &0.477&0.598&0.489&0.610&0.486 &0.610\\
 KECG (Orth. $W$) &0.481&0.601&0.499&0.605&0.497 &0.618\\
 KECG (Unit $W$) &\textbf{0.484}&\textbf{0.603}&\textbf{0.502}&\textbf{0.619}&\textbf{0.501} &\textbf{0.629}\\
 \hline
 AliNet &0.525&0.619&0.539&0.638&0.535 &0.645\\
 AliNet(Orth. $W$) &0.538&0.629&0.557&0.644&0.562 &0.657\\
 AliNet(Unit $W$) &\textbf{0.543}&\textbf{0.636}&\textbf{0.561}&\textbf{0.648}&\textbf{0.565} &\textbf{0.663}\\
\hline
\end{tabular}
}
\caption{Ablation experiment on complex methods\protect\footnotemark[5].}\label{table:2}
\end{table}

\noindent
(2) \textbf{Experiment on Complex GNNs}:
To further verify orthogonal is also necessary for complex methods,
we test orthogonal and unit constraint settings with MuGNN, KECG, and AliNet.
Originally, MuGNN and KECG adopt diagonal constraint while AliNet is unconstrained.
The experimental results are shown in Table \ref{table:2}.
It is obvious that both orthogonal and unit constraints improve the performances on all datasets compared to each method's original constraint setting.
The unit constraint is slightly better than orthogonal constraint.
This may be due to the fact that more transformation matrices are in complex methods, which make the orthogonal constraint slightly harder to optimize.

\footnotetext[5]{AliNet only releases part of the source code (w/o rel).}

In summary, we believe that \textbf{the transformation matrix $\bm{W}$ in GNNs should be constrained to be orthogonal} to ensure that the norms and the relative distances of entities remain unchanged after transformation.
Unit matrix is not only a special case of orthogonal but also the simplest implementation.
The experimental results prove that our conclusion is universal to both the simplest and complex GNNs-based methods.
Many existing GNNs-based methods could be further improved by adopting this setting.

\subsection{Why Advanced KG Embedding Not Work}
Many advanced KG embedding models are proposed and proven to be successful in link prediction task.
But a lot of them have very poor performances in entity alignment task as shown in Table \ref{table:KEM}.
For translation-based methods, they are at least $17\%$ worse than TransE, while for GNNs-based methods they are at least $3\%$ worse than GCN.
Why they are not working in with entity alignment?
To compare these KG embedding models clearly, we summarize their core functions in Table \ref{table:methods}.
From the table, we observe that all these advanced methods share one key idea:
transform universal entity embeddings into relation specific ones.
In particular, RGCN is a combination of GCN and TransR while KBAT references the ConvE and applies it to GAT.
However, in their original design, all of them do not put any constraint on their transformation matrix.
This violates our conclusion in Section \ref{4.3}.
Such unconstrained transformation destroys the \emph{shape similarity} and results in their poor performances for entity alignment task (Table \ref{table:KEM}).
\begin{table}[t]
\renewcommand\arraystretch{1.3}
\centering
\resizebox{1\linewidth}{!}{
\begin{tabular}{c|cc|cc|cc}
\hline
\multirow{2}{* }{Method} & \multicolumn{2}{c|}{$\rm{DBP_{ZH-EN}}$} & \multicolumn{2}{c|}{$\rm{DBP_{JA-EN}}$} & \multicolumn{2}{c}{$\rm{DBP_{FR-EN}}$}\\
& Hits@1 & MRR & Hits@1 & MRR & Hits@1 & MRR\\
\hline
 TransE \cite{DBLP:conf/nips/BordesUGWY13}&0.423&0.534&0.421&0.531&0.449& 0.568\\
 TransR* \cite{DBLP:conf/aaai/LinLSLZ15}&0.259&0.349&0.222&0.295&0.059& 0.116\\
 ConvE* \cite{DBLP:conf/aaai/DettmersMS018}&0.169&0.224&0.192&0.246&0.240& 0.316\\
 \hline
 GCN \cite{DBLP:conf/iclr/KipfW17}&\textbf{0.448}&\textbf{0.573}&\textbf{0.466}&\textbf{0.589}&\textbf{0.462}&\textbf{0.594}\\
 RGCN \cite{DBLP:conf/esws/SchlichtkrullKB18}& 0.419&0.505&0.424&0.517&0.431&0.561 \\
 KBAT \cite{DBLP:conf/acl/NathaniCSK19}& 0.323&0.381&0.311&0.363&0.307&0.362 \\
\hline
\end{tabular}
}
\caption{Performance of different KGs embedding models in entity alignment. * represents the result is taken from \citet{DBLP:conf/semweb/SunHHCGQ19}. Other results are produced by ourselves.}
\label{table:KEM}
\end{table}

\begin{table}[t]
\renewcommand\arraystretch{1.3}
\centering
\resizebox{1.0\linewidth}{!}{
\begin{tabular}{c|c|c}
Method & Embedding Function& $\varphi(\cdot)$\\
\hline
GCN \cite{DBLP:conf/iclr/KipfW17}& $ \sigma\left(\sum_{j\in{\mathcal N}_i}\frac1{\sqrt{d_id_j}}\varphi(\bm{h}_{e_j})\right) $ & $\bm{W}h$\\
TransR \cite{DBLP:conf/aaai/LinLSLZ15}& $\|\varphi(\bm{h},\bm{r})+\bm{r}-\varphi(\bm{t},\bm{r})\|$ & $\bm{W}_r\bm{h}$ \\
RGCN \cite{DBLP:conf/esws/SchlichtkrullKB18}&$\sigma(\sum_{r\in R}\sum_{j\in\mathcal N_i^r}\frac1{|N_i^r|}\varphi(\bm{h}_{e_j},\bm{r})+\bm{W}_0^l\bm{h}_{e_i}^l)$& $\bm{W}_r\bm{h}$\\
\hline
GAT \cite{DBLP:conf/iclr/VelickovicCCRLB18}& $ \sigma\left(\sum_{j\in{\mathcal N}_i}\alpha_{ij}\varphi(\bm{h}_{e_j})\right) $ & $\bm{W}\bm{h}$\\
ConvE \cite{DBLP:conf/aaai/DettmersMS018}& $ \sigma(\varphi(\bm{h},\bm{r}) \odot \bm{t}) $ & $\bm{W}vec([\bm{h}\|\bm{r}]\ast\omega)$\\
KBAT \cite{DBLP:conf/acl/NathaniCSK19}&$\sigma\left(\sum_{j\in{\mathcal N}_i}\sum_{k\in R_{ij}}\alpha_{ijk}\varphi(\bm{h}_{e_i},\bm{h}_{r_k},\bm{h}_{e_j})\right)$& $\bm{W}[\bm{h}\|\bm{r}\|\bm{t}]$\\
\hline
\end{tabular}
}
\caption{A summary of some representative KGs embedding models. $\|$ represents the concatenate operation. $\ast$ and $\omega$ represent the convolution operation and kernel. $d_i$ represents the degree of entity $e_i$.}\label{table:methods}
\end{table}

Theoretically, based on our conclusion in Section \ref{4.3}, if the transformation matrix in these advanced methods could comply to orthogonal, then the shape similarity would be reserved.
But such constraint is very difficult to adopt in practice.
For TransR and RGCN, because there are usually thousands of relations in KGs, constraining all the relational matrices is not feasible.
For ConvE and KBAT, the dimension of transformed embeddings must be kept consistent with that of input embeddings.
Otherwise, it will cause dimension mismatch in ConvE or dimension explosion when stacking multiple layers in KBAT.
Therefore, the transformation matrix of ConvE and KBAT cannot be a square matrix, let alone an orthogonal matrix.
But their successes in linked prediction bring one insight that constructing relation specific entity embedding is more effective in modeling relations,
compared to just assigning relation-based to entities.

\subsection{Key Criteria for Transformation Operation}
Therefore, the ideal transformation operation in entity alignment should satisfy the following two key criteria:

(1) \textbf{Relational Differentiation}:
Corresponding to different relation types, the operation could transform embedding of the same entity into different relational spaces.
\begin{equation}
    \varphi(\bm{h}_e,\bm{h}_{r_1}) \neq \varphi(\bm{h}_e,\bm{h}_{r_2}), \forall e\in E, \forall r_1,r_2\in R
\end{equation}

(2) \textbf{Dimensional Isometry}:
When two entities in the same KG are transformed into the same relational space, their norms and relative distance should be retained.
\begin{align}
    \|\bm{h}_e\| = \|\varphi(\bm{h}_e,\bm{h}_r)\| ,\; \forall &e\in E, \forall r\in R\\
    \bm{h}_{e_1}^T\bm{h}_{e_2} = \varphi(\bm{h}_{e_1},\bm{h}_r)^T\varphi(\bm{h}_{e_2},\bm{h}_r),&\; \forall e_1,e_2\in E, \forall r\in R
\end{align}

\section{The Proposed Method}
\label{sec:model}
In this section, we propose a novel GNNs-based method, \emph{Relational Reflection Entity Alignment} (RREA), which incorporates \emph{Relational Reflection Transformation} in GNNs to fulfill both relational differentiation and dimensional isometry criteria at the same time.

\subsection{Relational Reflection Transformation}
\begin{figure}[t]
\resizebox{0.9\linewidth}{!}{
  \centering
  \includegraphics[width=\linewidth]{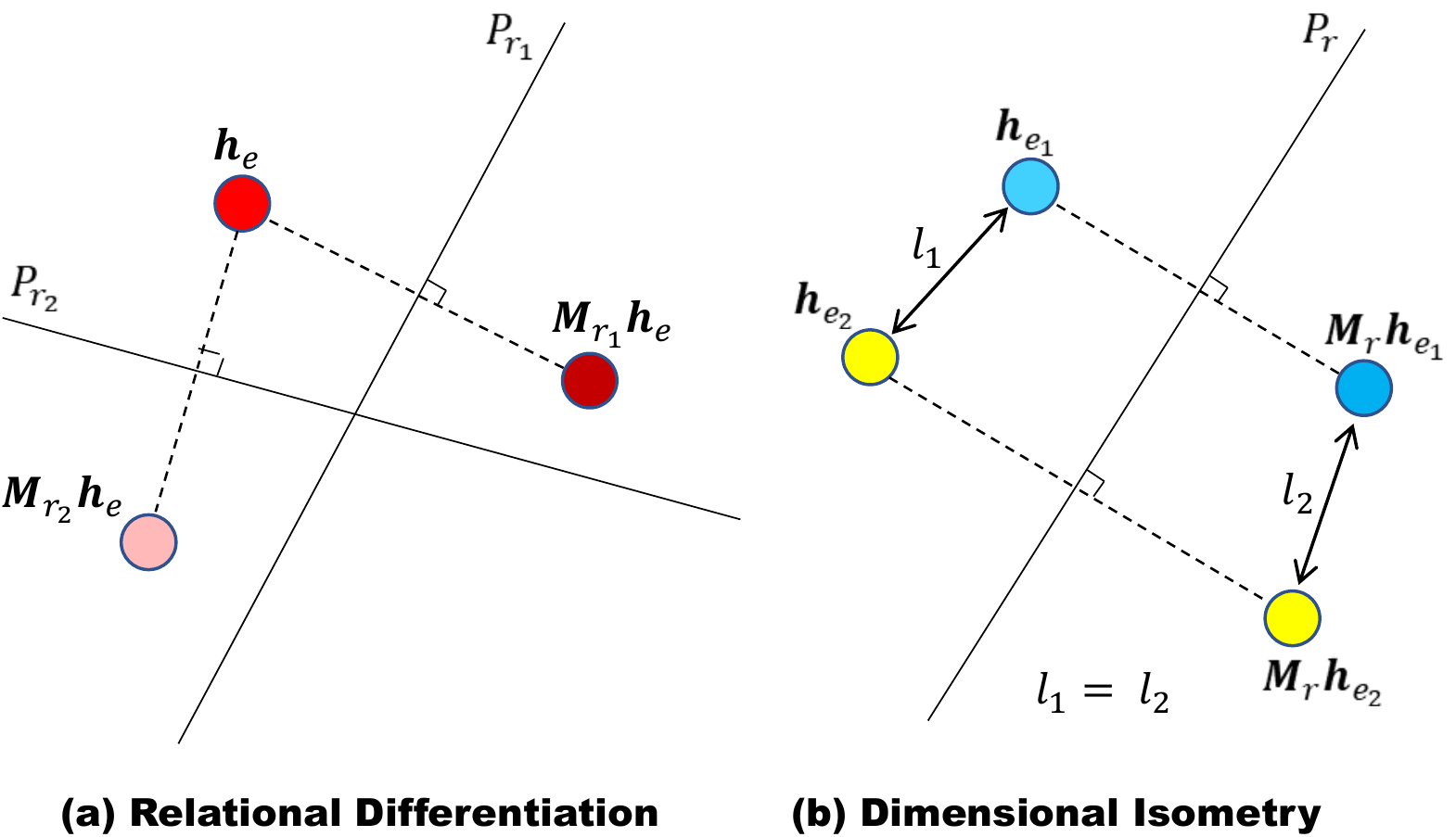}
  }
  \caption{The illustration of relational reflection operation.}\label{mirror}
\end{figure}

To meet the key criteria, we design a new transformation operation, \emph{Relational Reflection Transformation}.
Let relation embedding $\bm{h}_r$ be a normal vector, there is one and only one hyperplane $\bm{P}_r$ and only one corresponding reflection matrix $\bm{M}_r$ such that:
\begin{equation}
    \bm{M}_r= \bm{I}-2\bm{h}_r\bm{h}_r^T
\end{equation}
Here $\bm{h}_r$ should be normalized to ensure $\|\bm{h}_r\|_2=1$.
It is easy to derive that the reflection of entity embedding $\bm{h}_e$ along the relational hyperplane $\bm{P}_r$ can be computed by $\bm{M}_r\bm{h}_e$.
It is also easy to prove that $\bm{M}_r$ is orthogonal:
\begin{equation}
\begin{aligned}
    \bm{M}_r^T\bm{M}_r &= (\bm{I}-2\bm{h}_r\bm{h}_r^T)^T(\bm{I}-2\bm{h}_r\bm{h}_r^T)\\
             &= \bm{I} - 4\bm{h}_r\bm{h}_r^T + 4\bm{h}_r\bm{h}_r^T\bm{h}_r\bm{h}_r^T = \bm{I}
\end{aligned}
\end{equation}
Therefore, as long as \{$\bm{h}_{r_i} \neq \bm{h}_{r_j}, \forall r_i,r_j \in R$\}, our \emph{Relational Reflection Transformation} satisfies the two key criteria (illustrated as Figure \ref{mirror} (a) and (b)).

\subsection{Relational Reflection Entity Alignment}
In this section, we describe our proposed model \emph{Relational Reflection Entity Alignment} (RREA).
The inputs are two matrices: $\bm{H^e}\in\mathbb{R}^{\vert E\vert\times d}$ represents the entity embeddings and $\bm{H^r}\in\mathbb{R}^{\vert R\vert\times d}$ represents the relation embeddings.
Both $\bm{H^e}$ and $\bm{H^r}$ are randomly initialized by \emph{He\_initializer} \cite{DBLP:conf/iccv/HeZRS15}.
RREA consists of the following four major components:

\noindent
\textbf{Relational Reflection Aggregate Layer}:
The output feature of $e_i$ from the $l$-$th$ layer is obtained as follow:
\begin{equation}
\bm{h}_{e_i}^{l+1}={\rm ReLU}\left(\sum_{e_j\in \mathcal N_{e_i}^e}\sum_{r_k\in R_{ij}}\alpha_{ijk}^l \bm{M}_{r_k}\bm{h}_{e_j}^l\right)
\end{equation}
where $\mathcal N^e_{e_i}$ represents the neighboring entity set of $e_i$, $R_{ij}$ represents the set of relations between $e_i$ and $e_j$,
$\bm{M}_{r_k}\in\mathbb{R}^{d\times d}$ is the relational reflection matrix of $r_k$.
Compared with RGCN which assigns different $\bm{W}_r$ to different relations, the number of trainable parameters of relational reflection is much less because the degrees of freedom of $\bm{M}_r$ is only $d$ rather than $d^2$.
Similar to GAT, $\alpha^l_{ijk}$ represents the weight coefficient of $\bm{M}_{r_k}\bm{h}_{e_j}^l$ which is computed by the following equations:
\begin{equation}
  \beta_{ijk}^l = \bm{v}^T[\bm{h}_{e_i}^l\|\bm{M}_{r_k}\bm{h}_{e_j}^l\|\bm{h}_{r_k}]
\end{equation}
\begin{equation}
\alpha_{ijk}^l=\frac{exp(\beta_{ijk}^l)}{\sum_{e_j\in \mathcal N^e_{e_i}}\sum_{r_k\in R_{ij}}exp(\beta_{ijk}^l))}
\end{equation}
where $\bm{v}\in \mathbb{R}^{2d}$ is a trainable vector for calculating the weight coefficient.
To create a global-aware graph representation, we stack multiple layers of GNNs to capture multi-hop neighborhood information.
The embeddings from different layers are concatenated together to get the final output feature $\bm{h}^{out}_{e_i}$ of entity $e_i$:
\begin{equation}
 \bm{h}^{out}_{e_i} = [\ \bm{h}^{0}_{e_i}\| ...\| \bm{h}^{l}_{e_i}\ ]
\end{equation}
where $\bm{h}^{0}_{e_i}$ represents the initial embedding of $e_i$.

\noindent
\textbf{Dual-Aspect Embedding}:
Some recent studies \cite{yang2019aligning,DBLP:conf/wsdm/MaoWXLW20} believe that the entity embeddings generated by GNNs only contain the topological information, lack the relational information around entities.
Therefore, they concatenate the summation of the relation embeddings with entity embeddings to get dual-aspect embeddings.
In this paper, we adopt dual-aspect embeddings with the following equation:
\begin{equation}
\bm{h}^{Mul}_{e_i}=\left[\bm{h}^{out}_{e_i}\Big|\Big|\frac1{|\mathcal N^r_{e_i}|}\sum_{r_j\in \mathcal N^r_{e_i}}\bm{h}_{r_j}\right]
\end{equation}
where $\mathcal N^r_{e_i}$ represents the set of the relations around entity $e_i$.

\noindent
\textbf{Alignment Loss Function for Training}:
In order to make the equivalent entities close to each other in the unified vector space, we adopt the following triplet loss function:
\begin{equation}
\small
\begin{split}
L=\sum_{\substack{\left(e_i,e_j\right)\in P}}&{max}\left(dist\left(e_i,e_j\right)-dist\left(e'_i,e_j'\right)+\lambda,0\right)
\end{split}
\end{equation}
Here, $e'_i$ and $e'_j$ represent the negative pair of $e_i$ and $e_j$ which are generated by nearest neighbor sampling \cite{DBLP:conf/ijcai/SunHZQ18}.
In the training process, we take the same setting with GCN-Align \cite{DBLP:conf/emnlp/WangLLZ18} which uses Manhattan distance as the distance metric.
\begin{equation}
dist\left(e_i,e_j\right)=\left\| \bm{h}^{Mul}_{e_i}-\bm{h}^{Mul}_{e_j} \right\|_1
\end{equation}

\noindent
\textbf{CSLS Metric for Testing}:
We notice that \citet{DBLP:conf/iclr/LampleCRDJ18} propose \emph{Cross-domain Similarity Local Scaling} (CSLS) to solve the hubness problem existing in cross-lingual word embedding task.
Inspired by their study, we adopt CSLS as the distance metric during testing.

\subsection{Further Data Enhancement}

\noindent
\textbf{Semi-supervised Learning}:
In practice, the aligned seeds are often inadequate due to the high cost of manual annotations and the huge size of KG.
To expand training data, some recent studies \cite{DBLP:conf/ijcai/SunHZQ18,DBLP:conf/wsdm/MaoWXLW20} adopt iterative or bootstrapping strategies to build semi-supervised models.
In this paper, we use the iterative strategy proposed by MRAEA \cite{DBLP:conf/wsdm/MaoWXLW20} to generate semi-supervised data.

\noindent
\textbf{Unsupervised Textual Framework}:
The methods we have discussed before only focus on the structural information of KGs.
In some KGs, rich textual information are also available such as the entity names.
Therefore, some recent methods propose to combine textual information and structural information.
Among these methods, the unsupervised textual framework proposed by MRAEA \cite{DBLP:conf/wsdm/MaoWXLW20} does not require labeled data, which is more practical.
In this paper, we adopt the unsupervised textual framework from MRAEA.

\section{Experiments}
\label{exp}
In this section, we conduct a series of experiments on two public datasets to prove that our model not only outperforms all existing methods but also is robust.
The code is now available on GitHub\footnote{\url{https://github.com/MaoXinn/RREA}.}.

\subsection{Experiment Setting}
\textbf{Data Split and Metrics}:
Following previous studies, we randomly split $30\%$ of the pre-aligned entity pairs as training data and left the remaining data for testing.
The reported performance is the average of five independent training runs and the train/test datasets are shuffled in every round.
We also use \emph{Hits@k} and \emph{Mean Reciprocal Rank} (\emph{MRR}) to be the evaluation metrics as previous works.
\emph{Hits@k} represents the percentage of correctly aligned entities to the top-k potential entities.
The higher the \emph{Hits@k} and \emph{MRR}, the better the performance.

\noindent
\textbf{Hyper-parameters Selection}:
We select the hyper-parameters with the following candidate sets:
embedding dimension $d \in \{75, \\100, 150, 200\}$, margin $\lambda \in \{1.0, 2.0, 3.0, 4.0\}$, learning rate $\gamma \in \{0.001, 0.005, 0.01\}$,
GNN's depth $l \in \{1, 2, 3, 4\}$, dropout rate $\mu \in \{0.2, 0.3, 0.4, 0.5\}$.
For all of the datasets, we use a same config:
$d = 100$,  $\lambda=3$, $l = 2$, $\mu = 0.3$, $\gamma = 0.005$.
RMSprop is adopted to optimize the model and the number of epochs is set to $3,000$.

\subsection{Baselines}
As an emerging task, entity alignment attracts a lot of attention in a short time.
Many studies believe that the information of existing datasets is insufficient, so they try to introduce extra data into datasets.
For example, GMNN \cite{DBLP:conf/acl/XuWYFSWY19} and RDGCN \cite{DBLP:conf/ijcai/WuLF0Y019} use the name of entities as input features, BootEA \cite{DBLP:conf/ijcai/SunHZQ18} introduces semi-supervision to extend the datasets.
We believe that the introduction of extra data may lead to unfair comparisons between methods.
Therefore, we divide existing methods into three categories according to the data they use:

\begin{itemize}
  \item \textbf{Basic}:
This kind of methods only uses original structural data (i.e., triples) from the datasets:
JAPE \cite{DBLP:conf/semweb/SunHL17}, GCN-Align \cite{DBLP:conf/emnlp/WangLLZ18}, RSN \cite{DBLP:conf/icml/GuoSH19}, MuGNN \cite{DBLP:conf/acl/CaoLLLLC19}, TransEdge \cite{DBLP:conf/semweb/SunHHCGQ19}, AliNet \cite{sun2019knowledge} and MRAEA \cite{DBLP:conf/wsdm/MaoWXLW20}.
  \item \textbf{Semi-supervised}:
  This kind of methods introduces semi-supervision to generate extra structural data:
Boot-EA \cite{DBLP:conf/ijcai/SunHZQ18}, NAEA \cite{DBLP:conf/ijcai/ZhuZ0TG19}, TransEdge (semi), MRAEA (semi).
  \item \textbf{Textual}:
Besides the structural data, textual methods introduce entity names as additional input features:
GMNN \cite{DBLP:conf/acl/XuWYFSWY19}, RDGCN \cite{DBLP:conf/ijcai/WuLF0Y019}, HGCN \cite{DBLP:conf/emnlp/WuLFWZ19}, MRAEA (text) and DGMC \cite{DBLP:journals/corr/abs-2001-09621}.
\end{itemize}
Correspondingly, in order to make fair comparisons with all kinds of methods, our RREA also has three versions: RREA (basic), RREA (semi), and RREA (text).

\begin{table}[t]
\renewcommand\arraystretch{1.3}
\centering
\resizebox{1\linewidth}{!}{
\begin{tabular}{c|cc|cc|cc}
\hline
\multirow{2}{* }{Method} & \multicolumn{2}{c|}{$\rm{DBP_{ZH-EN}}$} & \multicolumn{2}{c|}{$\rm{DBP_{JA-EN}}$} & \multicolumn{2}{c}{$\rm{DBP_{FR-EN}}$}\\
& Hits@1 & MRR & Hits@1 & MRR & Hits@1 & MRR\\
\hline
 GMNN &0.679&0.785&0.740&0.872&0.894& 0.952\\
 RDGCN&0.708&0.846&0.767&0.895&0.886& 0.957\\
 HGCN&0.720&0.857&0.766&0.897&0.892& 0.961\\
 MRAEA&0.778&0.935&0.889&0.969&0.950& 0.984\\
 DGMC& 0.801&0.875&0.848&0.897&0.933&0.960 \\
\hline
 RREA& \textbf{0.822}&\textbf{0.964}&\textbf{0.918}&\textbf{0.978}&\textbf{0.963}&\textbf{0.992} \\
 Improv.& 2.62\%&3.10\%&3.26\%&0.93\%&1.37\%&0.81\%\\
\hline
\end{tabular}
}
\caption{Experimental results of textual methods.}
\label{table:textual}
\end{table}

\begin{table*}[!t]
\renewcommand\arraystretch{1.1}
\begin{center}
\resizebox{\textwidth}{!}{
\begin{tabular}{c|c|ccc|ccc|ccc|ccc|ccc}
  \multicolumn{2}{c|}{\multirow{2}{*}{Method}} & \multicolumn{3}{c|}{$\rm{DBP_{ZH-EN}}$} & \multicolumn{3}{c|}{$\rm{DBP_{JA-EN}}$} & \multicolumn{3}{c|}{$\rm{DBP_{FR-EN}}$}& \multicolumn{3}{c|}{$\rm{DWY_{WD}}$}& \multicolumn{3}{c}{$\rm{DWY_{YG}}$}  \\
  \multicolumn{2}{c|}{} & H@1 & H@10 & MRR & H@1 & H@10 & MRR & H@1 & H@10 & MRR & H@1 & H@10 & MRR & H@1 & H@10 & MRR\\
  \hline
  \multirow{10}*{Basic} 
  & $\triangle$JAPE & 0.411 & 0.744 & 0.490 & 0.362 & 0.685 & 0.476 & 0.323 & 0.666 & 0.430 &0.318& 0.589&0.411&0.236&0.484&0.320\\
  & GCN-Align & 0.412 & 0.743 & 0.549 & 0.399 & 0.744 & 0.546 & 0.372 & 0.744 & 0.532 & 0.506& 0.772&0.600&0.597&0.838& 0.682\\
  & $\triangle$RSN&0.508&0.745&0.591&0.507&0.737&0.590&0.516&0.768&0.605&0.607&0.793& 0.673& 0.689& 0.878&0.756\\
  & MuGNN & 0.494 & 0.844 & 0.611 & 0.501 & 0.857 & 0.621 & 0.495 & 0.870  &0.621 &0.616&0.897& 0.714& 0.741&0.937&0.810\\
  & KECG&0.477&0.835&0.598& 0.489&0.844& 0.610&0.486&0.851& 0.610& 0.632& 0.899& 0.726&0.728 &0.915& 0.795\\
  & AliNet&0.539&0.826&0.628&0.549&0.831&0.645&0.552&0.852&0.657&0.690& 0.908& 0.766&0.786& 0.943& 0.841\\
  & $\triangle$TransEdge&0.659&0.903&0.748&0.646&0.907&0.741&0.649&0.921&0.746&0.692&0.898&0.770&0.726&0.909&0.792\\
  & MRAEA &0.638&0.886&0.736&0.646&0.891&0.735&0.666&0.912&0.765&-&-&-&-&-&-\\

  \cline{2-17}
  & $\rm{RREA}$ &\textbf{0.715}&\textbf{0.929}&\textbf{0.794}&\textbf{0.713}&\textbf{0.933}&\textbf{0.793}&\textbf{0.739}&\textbf{0.946}&\textbf{0.816}&\textbf{0.753}&\textbf{0.945}&\textbf{0.824}&\textbf{0.839}&\textbf{0.968}&\textbf{0.887}\\
  & Improv.&8.49\%&2.88\%&6.15\%&10.4\%&2.87\%&7.02\%&10.9\%&3.73\%&6.67\%&9.13\%&4.07\%&7.57\%&6.74\%&2.65\%&5.47\%\\
  \hline
  \multirow{6}*{Semi}
  & $\triangle$BootEA & 0.629 & 0.847 & 0.703 & 0.622 & 0.853 & 0.701 & 0.653 & 0.874 & 0.731 & 0.747&0.898&0.801&0.761&0.894&0.808\\
  & NAEA & 0.650 & 0.867 & 0.720 & 0.641 & 0.872 & 0.718 & 0.673 & 0.894 & 0.752  &0.767&0.917& 0.817&0.778&0.912&0.821\\
  & $\triangle$TransEdge&0.735&0.919&0.801&0.719&0.932&0.795&0.710&0.941&0.796&0.788&0.938&0.824&0.792&0.936&0.832\\
  & MRAEA &0.757&0.930&0.827&0.758&0.934&0.826&0.781&0.948&0.849&-&-&-&-&-&-\\
  \cline{2-17}
  & RREA &\textbf{0.801}&\textbf{0.948}&\textbf{0.857}&\textbf{0.802}&\textbf{0.952}&\textbf{0.858}&\textbf{0.827}&\textbf{0.966}&\textbf{0.881}&\textbf{0.854}&\textbf{0.966}&\textbf{0.877}&\textbf{0.874}&\textbf{0.976}&\textbf{0.913}\\
  & Improv.&5.81\%&1.94\%&3.63\%&5.80\%&1.93\%&3.87\%&5.89\%&1.90\%&3.77\%&8.37\%&2.99\%&6.43\%&10.3\%&4.27\%&9.73\%\\
    \hline
\end{tabular}
}
\caption{Experimental results of basic and semi-supervised methods. "Improv." represents the percentage increase compared with SOTA. $\triangle$ represents translation-based methods.}
\label{table:res1}
\end{center}
\end{table*}

\begin{figure*}[t]
  \includegraphics[width=\textwidth]{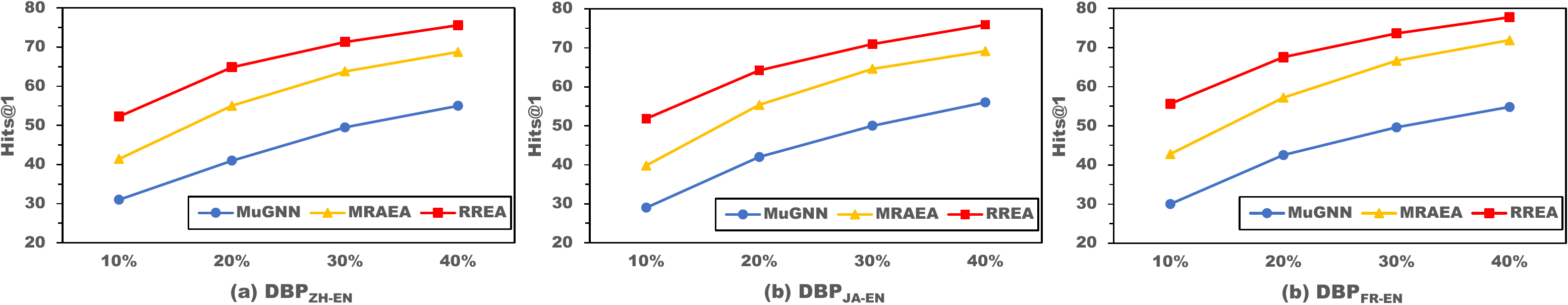}\\
  \caption{\emph{Hits@1} performances of different pre-aligned ratios on DBP15K.}
  \label{line-chart2}
\end{figure*}

\subsection{Main Results and Ablation Studies}
\textbf{RREA vs. Basic and Semi-supervised Methods}.
Table \ref{table:res1} shows the performance comparisons for basic and semi-supervised methods.
Obviously, the performances of our model are consistently ranked as the best over all competing basic methods and semi-supervised methods on all the evaluation metrics.
Especially, compared with the state-of-the-art methods TransEdge and MRAEA, RREA (basic) exceeds by at least $6\%$ on \emph{Hits@1} and RREA(semi) exceeds by more than $5\%$ on \emph{Hits@1} respectively.
The main reason is that our reflection transformation builds relation specific embeddings for entities which could capture the relation information better.
In addition, it is clear that semi-supervision could significantly improve the performances of all the methods on all datasets.
Compared to RREA (basic), RREA (semi) iteratively generates extra training data via semi-supervision which improves the performance by an average of $6\%$ on \emph{Hits@1}.
In summary, RREA breaks the performance ceiling of purely structural-based entity alignment methods, which proves that our designs are effective.

\noindent
\textbf{RREA vs. Textual Methods}.
Since all the datasets of $\rm DWY100K$ are sampled from English KGs, the textual information is highly similar.
Therefore, we only conduct the experiments of textual methods on $\rm DBP15K$.
Table \ref{table:textual} shows the results of the compared methods.
Our model beats MRAEA and achieves the best on all datasets.
Since we use the unsupervised textual framework proposed by MRAEA, the performance improvement is totally contributed by the better modelling of structural data.
Compared with other supervised models (e.g., DGMC, GMNN), RREA (text) even achieves better performance while using the same datasets.

We observe that the performance gap of textual methods between different datasets is far bigger than that of structural methods.
All methods perform much better in French than in the other two languages.
That is because the difference between French and English is much smaller than the others.
So French words are easier to be mapped to English by cross-lingual word embedding or machine translation.
In addition, although the performances of textual methods are significantly better than that of structural methods, the structural methods are more universal in practice.
Because the current datasets are all sampled from Wikipedia, the textual information such as entity names is too simple for Google translation or cross-lingual embedding whose training corpus are also sampled from Wikipedia.
In reality, textual information often is not available, or it is very hard to get a high quality translation.
Therefore, we believe that the textual methods should be compared separately in studies, rather than with the structural methods together.

\begin{table}[t]
\renewcommand\arraystretch{1.5}
\centering
\resizebox{1.0\linewidth}{!}{
\begin{tabular}{l|cc|cc|cc}
\hline
\multirow{2}{* }{Method} & \multicolumn{2}{c|}{$\rm{DBP_{ZH-EN}}$} & \multicolumn{2}{c|}{$\rm{DBP_{JA-EN}}$} & \multicolumn{2}{c}{$\rm{DBP_{FR-EN}}$}\\
& Hits@1 & MRR & Hits@1 & MRR & Hits@1 & MRR\\
\hline
 GCN-Align &$.449_{\pm .002}$&$.574_{\pm .002}$&$.464_{\pm .003}$&$.588_{\pm .002}$&$.463_{\pm .004}$&$ .596_{\pm .005}$\\
 \;\;+CSLS&$.487_{\pm .002}$&$.601_{\pm .002}$&$.507_{\pm .003}$&$.620_{\pm .002}$&$.503_{\pm .004}$&$ .487_{\pm .005}$\\
 \;\;+Rel. Refl.&$.631_{\pm .002}$&$.724_{\pm .002}$&$.644_{\pm .005}$&$.738_{\pm .003}$&$.667_{\pm .004}$&$ .761_{\pm .003}$\\
 \;\;+D-A Emb. &$.715_{\pm .002}$&$.794_{\pm .001}$&$.713_{\pm .001}$&$.793_{\pm .002}$&$.739_{\pm .002}$&$ .816_{\pm .001}$\\
\hline
\end{tabular}
}
\caption{Ablation experiment of RREA (basic) on DBP15K.}
\label{table:ablation}
\end{table}

\noindent
\textbf{Ablation Studies}.
In the above experiments, we have shown the overall success of RREA.
In this part, we want to demonstrate the effectiveness of each component in RREA (basic).
As mentioned in Section \ref{sec:model}, RREA (basic) has three designs compared with GCN-Align: (1) \emph{Cross-domain Similarity Local Scaling}; (2) \emph{Relational Reflection Aggregate Layer}; (3) \emph{Dual-Aspect Embedding}.
Starting from GCN-Align baseline, we gradually adopt these components and report the results with ${\rm Means}_{\pm {\rm stds.}}$ in Table \ref{table:ablation}.
Obviously, all of these three designs significantly improve performance.
Compared to GCN-Align, the introduction of CSLS improve performance by about $4\%$.
That shows the high correlation between entity alignment task and cross-lingual word embedding.
Adding \emph{Relational Reflection Aggregate Layer} and \emph{Dual-Aspect Embedding} to the model further brings about $15\%$ and $7\%$ improvement on $Hits@1$ respectively.
This means that both of the two designs introduce unique information into the model.
These ablation experiments show that our designs are meaningful and bring significant improvements.

\subsection{Robustness Analysis}
\textbf{Robustness on Pre-aligned Ratio.}
Generally speaking, building pre-aligned seeds is a high resource-consuming operation.
Especially when practicing in the real-world, the KGs usually have millions of entities, relations, and triples.
Therefore, we hope that the model could perform well in a lower pre-aligned resource situation.
To investigate the robustness of RREA in different pre-aligned ratios, we compare the performance of three GNN-based methods on $\rm{DBP15K}$ (MuGNN, MRAEA, and RREA (basic)) with different ratios of pre-aligned pairs.
Figure \ref{line-chart2} reports their performance when reserving $10\%$ to $40\%$ of pre-aligned pairs as training data on each of three cross-lingual datasets.
Obviously, RREA significantly outperforms compared methods in all pre-aligned ratios of training data.
With only $10\%$ pre-aligned pairs, RREA (basic) still achieves more than $52\%$ \emph{Hits@1} on $\rm DBP15K$, which even better than the performance of MuGNN in $40\%$ pre-aligned ratio.

\begin{figure}[t]
  \includegraphics[width=\linewidth]{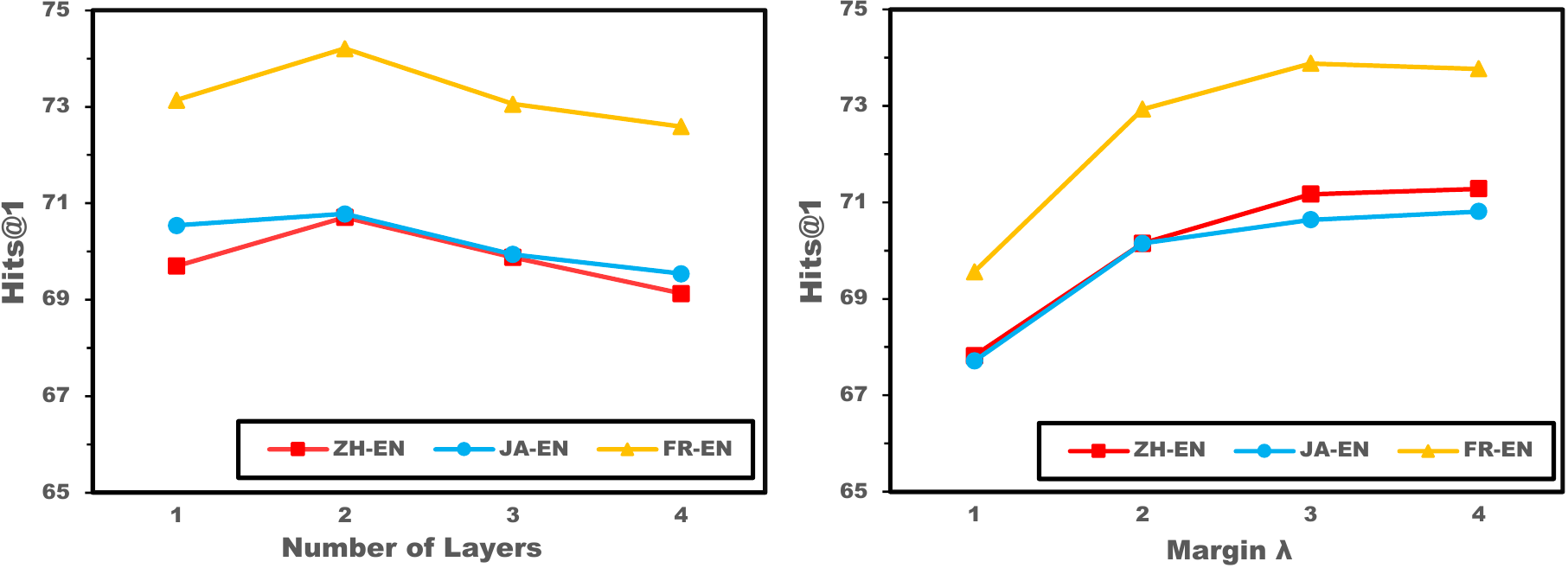}\\
  \caption{Hyper-parameter studies on DBP15K.}
  \label{line-chart}
\end{figure}
\noindent
\textbf{Robustness on Hyper-parameter.}
In order to investigate the robustness of RREA on hyper-parameters, we evaluate the performance on $\rm DBP15K$ varying the number of layer $l$  and the margin $\lambda$
while keeping the other hyper-parameters consistent with the default setting.
The experiment results are shown in Figure \ref{line-chart}.
For layer depth $l$, RREA with $2$ layers achieves the best performance on all datasets.
When stacking more layers, the performance begins to decrease slightly.
Stacking more layers only results in slower speed, not better performance.
For margin $\lambda$, when $\lambda$ is set to $2.0$$\sim$$4.0$, the performance gap is less than $1\%$.
In general, the impact of $l$ and $\lambda$ on performance is limited and the model is relatively stable during the varying of hyper-parameters.

\section{Conclusions}
In this paper, we raise the counter-intuitive phenomena in entity alignment, which are neglected by previous studies.
By abstracting existing entity alignment methods into a unified framework, we successfully explain the questions and derive two key criteria for transformation operation in entity alignment:
relational differentiation and dimensional isometry.
Inspired by these findings, we propose a novel GNNs-based method, \emph{Relational Reflection Entity Alignment} (RREA) which leverages a new transformation operation called relational reflection.
The experimental results show that our model is ranked consistently as the best across all real-world datasets and outperforms the state-of-the-art method more than $5.8$\% on \emph{Hits@1}.

\bibliographystyle{ACM-Reference-Format}
\bibliography{acmart}
\appendix

\end{document}